\documentclass{aa}
\usepackage{natbib,psfig,graphicx}
\begin{document}


\title{Large scale diffuse light in the Coma cluster: a multi-scale approach}

\offprints{C. Adami \email{christophe.adami@oamp.fr}}

\author{
C. Adami\inst{1} \and E. Slezak\inst{2} \and F. Durret\inst{3} \and
C.J. Conselice\inst{4} \and J.C. Cuillandre\inst{5} \and
J.S. Gallagher \inst{6} \and A. Mazure\inst{1} \and R.~Pell\'o\inst{7} 
\and J.P. Picat\inst{7} \and M.P. Ulmer\inst{8} 
}

\institute{
LAM, Traverse du Siphon, 13012 Marseille, France
\and
Observatoire de la C\^ote d'Azur, B.P. 4229, 06304 Nice Cedex 4, France
\and
Institut d'Astrophysique de Paris, CNRS, Universit\'e Pierre et Marie Curie,
98bis Bd Arago, F-75014 Paris, France 
\and
Department of Astronomy, Caltech, MS 105-24, Pasadena CA 91125, USA
\and
Canada-France-Hawaii Telescope Corporation, 65-1238 Mamalahoa Highway,
Kamuela, HI 96743 
\and
University of Wisconsin, Department of Astronomy, 475 N. Charter St., 
Madison, WI 53706, USA
\and
Observatoire Midi-Pyr\'en\'ees, 14 Av. Edouard Belin, 31400 Toulouse,
France
\and
NWU, Dearborn Observatory, 2131 Sheridan, 60208-2900 Evanston, USA
}

\date{Accepted . Received ; Draft printed: \today}

\authorrunning{Adami et al.}
\titlerunning{Large scale diffuse light in the Coma cluster: a
multi-scale approach}

\abstract{ We have obtained wide field images of the Coma cluster in
the B, V, R and I bands with the CFH12K camera at CFHT. To
search for large scale diffuse emission, we have applied to these
images an iterative multiscale wavelet analysis and reconstruction
technique which made it possible to model all the sources (stars and galaxies)
and subtract them from the original images. We found various
concentrations of diffuse emission present in the central zone around
the central galaxies NGC~4874 and NGC~4889. We characterize the
positions, sizes and colors of these concentrations. Some sources do
not seem to have strong star formation, while one
probably exhibits spiral-like
colors. One possible origin for the star forming diffuse emission
sources is that in the region of the two main galaxies NGC~4874 and
NGC~4889 spiral galaxies have recently been disrupted and star
formation is still active in the dispersed material. We also use the
characteristics of the sources of diffuse emission to trace the
cluster dynamics. A scenario in which the group around NGC~4874 is
moving north is consistent with our data.

\keywords{galaxies: clusters: individual: Coma cluster (Abell 1656)}
}

\maketitle

\section{Introduction}

Diffuse emission in clusters of galaxies may put strong constraints
on cluster formation
processes, as well as on the mechanisms that might account for
the enrichment of the intergalactic medium such as galaxy harassment
or stripping.

The existence of intergalactic matter visible at optical wavelengths
in the Coma cluster was already reported by Zwicky (1951), and several
searches based on photographic plate material were undertaken in the
1970's on Abell~2670 (Oemler 1973) and Coma (Thuan \& Kormendy
1977). Interestingly, Thuan \& Kormendy (1977) already noted that the
diffuse light appeared to be bluer than the galaxies, a fact
consistent with the idea that it is made of stars tidally stripped
from galaxies, which are usually bluer at large radii. Later on, CCDs
made it possible to better model and subtract galaxies (including the
giant central galaxy) from images to obtain the diffuse light
distribution after applying various treatments (Gudehus 1989, Uson et
al. 1991, V\'\i lchez-G\'omez et al. 1994, Gonzalez et al. 2000).

Besides, there has also been an intensive search for various types of
stars which could belong to an intergalactic population, such as
planetary nebulae in Virgo (Arnaboldi et al. 1996, 2003, Feldmeier et
al. 1998) and in Fornax (Theuns \& Warren 1997), red giants in Virgo
(Durrell et al. 2002) or the progenitors of two supernovae in
Abell~403 and Abell~2122/4 (Gal-Yam et al. 2003). A global result is
that these stars could contribute up to 5-20\% of the cluster total
luminosity, thus making this component an important one when analyzing
the optical properties of clusters.

The idea that diffuse light in clusters is probably due to tidal
disruptions of galaxies has recently been confirmed numerically by
Calc\'aneo-Rold\'an et al. (2000). These authors have applied
numerical simulations to show that the arc of diffuse light detected
in the Centaurus cluster is probably the debris of a tidally disrupted
spiral galaxy. With the purpose of observing more tidal debris in
clusters, deep imaging surveys have recently been initiated which show
evidence for plumes and arclike structures (Feldmeier et al. 2002,
2003, Mihos 2003 or Gregg $\&$ West 1998).

We have undertaken an extensive analysis of the Coma cluster, based on
deep multiband imaging (B, V, R and I filters) of a region of
42$\times$50~arcmin$^2$ which will be presented in subsequent
papers. To search for diffuse light features
we used a wavelet analysis and reconstruction of the image to recover
and subtract all the objects (stars and galaxies) from the raw
image. We will compare the diffuse
features revealed by our treatment to those found by Gregg \& West
(1998). Note that Trentham \& Mobasher (1998) also reported the
presence of a number of low surface brightness features in Coma. 

The data will be briefly presented in Sect.~\ref{data} and our method in
Sect.~\ref{analysis}. Results will be presented in
Sects.~\ref{resultsR} and \ref{resultsBV}. A possible scenario to
explain our results will be presented and discussed in
Sect.~\ref{scenario} and conclusions are summarized in
Sect.~\ref{conclusions}.

\section{The data}
\label{data}

The full set of data at our disposal is described in a companion paper
(Savine et al. in preparation). We list however the main points. 

Images were obtained at the 3.6m CFH telescope in the spring of 1999 and
of 2000 with the CFH12K CCD
camera of 12365$\times$8143 pixels$^2$, with a scale of 0.206
arcsec/pixel  in the CFH12K B, V, R
and I filters (close to Johnson filters but not identical), for
an area of about 
42$\times$50~arcmin$^2$ centered on
the dominant galaxies of the Coma cluster, NGC~4874 and NGC~4889. This area
was covered by two CFH12K fields which overlap by $\sim$7' in the
North/South direction. Total exposure times were 7200s in 5 
exposures for the B band, 5040s in 7 exposures for the V band 
and the North field, 2400s in 5 exposures for the V band and the South
field, 3300s in 6 exposures for the R band and the South field and
3600s in 5 exposures for the R band and the North field. Superflat fields
were used (combinations of several empty exposures delivered as a final
product by the CFHT pipeline). The B, V, R and I data are complete down to
R$\sim$24. Final image qualities range from 0.85 to 1 arcsec.

In this paper we choose to use only B, V and R band data for the whole
field of view. We discarded the I band data which present some
residual artefacts from the sky subtraction. We analyzed the R band
data, which are the deepest, in the whole field of view and, the
B and V band data only in the regions of interest, where diffuse
emission is detected in the R band.

\section{Data analysis}
\label{analysis}

\subsection{Method}

\begin{figure}

\caption{
A two step methodology is required to detect and restore faint features 
close to bright ones with wavelets. All images are 1.75'$\times$1.75'
in the R band.
(a) original image with objects of different shapes and surface
brightnesses partly superimposed on the halo of a very bright star
in the neighborhood;
(b) image restored from a thresholded wavelet transform selecting
coefficients in the scale range $2^0-2^9$pixels in radius with an
absolute value larger than 3 times the rms scatter expected for pure
noise -- some faint features have been missed;
(c) image obtained by subtracting the denoised image from the
original one, showing the faint features at various scales which
have been missed at the first step as well as some annular restoration
artefacts due to local saturation effects;
(d) denoised image of the residual image computed from the subset
of its wavelet coefficients in the scale range $2^0$-- $2^7$ pixels found
significant with respect to the same thresholds as in the first
step. 
}
\label{fig:wave1}
\end{figure}
%
%
\begin{figure}
\caption{
Final products of the multiscale image processing involved when
removing in a non-parametric way objects within a given scale range
from an image.  Images are 1.75'$\times$1.75' in the R band.
(a) denoised image of the whole set of small scale structures
in the original image as obtained by adding together the two
images restored from the significant wavelet coefficients of
the raw and residual images;
(b) difference between the original image and the image
of the objects, showing that, except from a few annular artefacts
due to local saturation, only noise and very large scale
components are present in this second residual image.
}
\label{fig:wave2}
\end{figure}

Our aim is to detect large scale faint diffuse light emission sources
in unknown locations of an astronomical image in the presence of
structures (stars and galaxies) of different scales. The usual way is
to remove these luminous objects. An attractive approach is to perform
a multiscale analysis of the image, assuming that the spatial
extension of the diffuse sources is in most cases larger than the
typical star or galaxy scales and just keep the large scale components. 
No prior identification or
modeling of the structures to be subtracted are required as was done
for instance by V\'\i lchez-G\'omez et al. (1994). The efficiency of
the multiscale analysis approach depends of course on the initial
assumption for the separation criterion between different structures,
but in any case the noise reduction due to the method should make
easier the retrieval of diffuse light structures with scales comparable
to those of stars and galaxies.

However, large scale saturated star haloes remain a problem. Such
features are so extended that removing completely their contributions
would require using very large scales in the treatment described
below, and this would also remove large scale diffuse light
sources. When possible, we therefore excluded the regions occupied by
these very bright stars (two out of three).

To obtain a denoised image of the objects smaller than a maximum size
we applied the method described in Bijaoui \& Ru\'e (1995) and Durret
et al. (2002) with
some modifications that improved the quality of the final result. We will
only recall the salient points of the treatment. The
method involves a thresholding of the data in the wavelet space in order
to remove the noise locally without smoothing the astrophysical
signal. This is explained by the fact that, since wavelets are functions with
a mean value equal to zero, 
a continuous object in real space appears at any scale in the wavelet
space as a region of positive wavelet coefficients surrounded by a
border of negative coefficients. It appears that the characteristics
of the object, which can be computed from the subset of coefficients that are
statistically significant with respect to the noise rms scatter, are
better estimated when these negative values are also considered, especially for
bright objects. 

Note that wavelet coefficients are considered to be significant at a
given scale when their absolute magnitude exceeds 3 times the RMS
fluctuation expected for the wavelet coefficients of white noise at
this same scale.  Larger thresholds are introduced for the first
(smallest) and second scales (4.0 and 3.5 times the rms fluctuations
respectively) in order to remove most of the spurious unavoidable
false alarms at these scales.  Hence, we first decided to select not
only positive but also negative significant wavelet coefficients when
thresholding the wavelet transform of the image.

Second, negative
coefficients for one object locally decrease the positive coefficients
linked to any object close enough to this first object. When detecting
real structures in a noisy image by means of their corresponding
domains in the wavelet space, faint objects close to a bright object
can therefore easily be missed.  This can be a serious problem because
we have in our catalog faint objects distributed in an area with
very bright objects in the center (as expected for a cluster
of galaxies).
Consequently, in order to carry out a more complete survey of objects
within a given size range, we decided to perform the analysis
twice. From the first subset of significant wavelet coefficients, a
positive image in real space is restored using an iterative conjugate
gradient algorithm with a regularization constraint (Appendix A in
Ru\'e \& Bijaoui 1997). This image is subtracted from the original one and
the wavelet transform of the residual is thresholded with similar
thresholds as in the first iteration. Applying the restoration
algorithm to the second subset of coefficients enables one to get
another image including the previously hidden significant features 
(see Fig.~1d).

In this way, adding together the two restored images with a positivity
constraint for the result (cf. the subtraction step) enables one to
obtain a more adequate map of the small scale structures within
the image. This last map can then be subtracted from the original
image, leading to a new residual image where large scale emission
components can be quite easily searched for.
Hence, the final products of this process are an image of what we call
the objects, i.e. the signal with a characteristic scale smaller than
the maximal scale used, and a residual image exhibiting mostly
features with a characteristic scale larger than this maximum value
(see Fig. 2).

\subsection{Application}

The images to be examined contain the two dominant galaxies of the
Coma cluster of galaxies as well as three very bright stars in 
the northern field. To handle these large objects we decided to
perform the first wavelet transform of our analysis using ten scales
(from a minimum scale $a=2^{0}=1$ pixel to a maximum scale
$a=2^{9}=512$ pixels). Features with a characteristic size larger than
512 pixels are therefore preserved in the first residual image, as
well as a number of previously smaller hidden features.

The map of these
remaining small scale structures has been obtained by means of a
wavelet transform of this first residual with a computation implying
only the first six, seven or eight scales of the dyadic $a=2^j$ scheme
for checking purposes.  A smaller set of scales than before is
prefered so as to preserve in the final residual image most of the diffuse
emission regions. For example, using the scale combination 10--6 (a
multiscale examination with 10 and 6 scales at the first and second
iteration, respectively), only the smallest diffuse sources appear on
the object image, along with galaxies of various sizes, whereas large
emission regions like the haloes of dominant galaxies or extended
diffuse light appear on the final residual image. The scale
combinations 10--7 and 10--8 remove from the residual image the
sources that are not very extended (i.e. with a typical diameter
smaller than $2^7=128$ and $2^8=256$ pixels, respectively). Scale  
combinations 10--6, 10--7 and 10--8 will
roughly preserve objects larger than 40$\arcsec$ (about 3 times the
characteristic scale), 1.5' and 3' respectively.

The various steps of our analysis are shown for an example region in
Figs.~\ref{fig:wave1} and \ref{fig:wave2}. This region corresponds to
source 3 as described in the following.

\section{Results for the R band data}
\label{resultsR}

\subsection{10-7 and 10-8 combination scales for the whole field of 
view: general overview}

We first describe in this section the results of the treatment for the
whole field of view using two scale combinations: 10-7 and 10-8. We
analyzed separately the northern and southern mosaics. These mosaics
overlap in the NGC~4874 and NGC~4889 area. We show the results with
the 10-7 scale combination for the southern mosaic in
Fig.~\ref{fig:diffusRsud} (reconstructed and residual images) and for
the northern mosaic in Fig.~\ref{fig:diffusRnord} (residual image).

\begin{figure}
\caption[]{Southern mosaic in the R band. 
Upper figure: reconstructed image of objects computed with the scale
combination 10-7. Lower figure: residual image computed with the scale
combination 10-7. We clearly see strong diffuse light emission towards
the top of the lower image.  The hemispheric structure corresponding
to the light scattered by the instrument is visible in the bottom part
of the lower image. North is up and East is left.}
\label{fig:diffusRsud}
\end{figure}

\begin{figure}
\caption[]{Residual image of Northern mosaic computed with the scale
combination 10-7 in the R band. In the bottom part of the
image we clearly see diffuse emission in the region of galaxies
NGC~4874 and 4889.  Just above this diffuse emission, we can see the
halo corresponding to the light of the bright star scattered by the
instrument. The rectangle at the upper right was not considered
because it contained two bright stars. North is up and East is left.}
\label{fig:diffusRnord}
\end{figure}

We removed from the northern mosaic the areas around two bright stars
located in the northwest part of the mosaic. Instrumental scattered
light from these objects was so intense that it affected our
treatment quite adversely. We only kept the area around the brightest star 
(just north
of NGC~4874) because the main signal we detected was close to these
coordinates.

In Figs.~\ref{fig:diffusRsud} and \ref{fig:diffusRnord} we see a
significant emission in the residual images around the dominant
galaxies. This emission appears with both scales (10-7 and 10-8), and
shows various features corresponding to different characteristic
emission scales. This emission is detected in both northern and
southern mosaic images, assuring us that the detection is not due to
an instrumental effect such as scattered light, since the same regions
on the sky are not located at the same place on the camera in the
northern and southern fields.

We detected a large diffuse emission around the bright star in the
northern mosaic (instrumental effect impossible to remove with our
method without removing smaller scale diffuse light sources). We will
therefore use preferentially the southern data to study the diffuse
emission in the dominant galaxy area because these data are much less
affected by bright stars. The diffuse light coming from the star
located in the northern mosaic is still visible in the southern mosaic,
but our treatment interprets this contribution as an object (it is
visible in the object image) and removes it from the residual
image. Contamination by this star in the southern mosaic is therefore
negligible.

We detected in the southern field a nearly circular structure centered
on the field center and mainly visible in the south part of the
southern field. This is the light scattered by the instrument itself
(see the CFHT database). This structure is annular with a central
emission right in the middle of the field (it is indicated as
``instrumental scattered light'' in Fig.~\ref{fig:diffusRsud}). It is
not detected in the northern field because of the areas removed around
the two bright stars and because of the large scale halo of the
brightest star of the field.

\subsection{10-6, 10-7 and 10-8 combination scales for the dominant 
galaxy cluster region}

We study here a region of 14$\times$14~arcmin$^2$ around NGC~4874 and
NGC~4889.  In order to detect smaller scale structures in this region
we also use a third scale combination: 10-6. The results are shown in
Figs. \ref{fig:2310-8}-\ref{fig:2310-63D}.

These figures show the residual images with several characteristic
positions: dominant galaxy positions, center of the bulk of the X-ray emission
from the cluster (Neumann et al. 2003), center of the faint-galaxy
density map (Biviano et al. 1996), dynamical center of the galaxy
group associated to NGC~4874 (Gurzadyan $\&$ Mazure 2001) and a
secondary X-ray peak (Finoguenov 2004, private communication).
Fig.~\ref{fig:2310-63D} shows two 3D maps with two view angles.

\begin{figure}
\caption[]{ R band residual image of the area containing the two dominant
galaxies computed with the 10-8 scale combination. Additional
locations are shown: dynamical center of the NGC~4874 group from
Gurzadyan $\&$ Mazure (2001)(cross), faint galaxy center from Biviano
et al. (1996)(square) and global X-ray center (open
circle). Black-shaded areas show the diffuse light emission maximum
(source~1). The two filled circles are the positions of the two
cluster dominant galaxies. The image size is 14$\times$14 arcmin$^2$, 
as in the three following figures.}
\label{fig:2310-8}
\end{figure}


\begin{figure}
\caption[]{ R band residual image of the area containing the two
dominant galaxies computed with the 10-7 scale combination. Additional
locations are shown as in Fig.~\ref{fig:2310-8}, together with the
secondary X-ray peak from Finoguenov (2004, private
communication)(right filled circle). Black shaded areas show the
diffuse light emission maxima, with sources 2 (left) and 3 (right).}
\label{fig:2310-7}
\end{figure}


\begin{figure}
\caption[]{R band images. Upper figure: object image of the area
containing the two dominant galaxies computed with the 10-6 scale
combination. Lower figure: residual image of the area containing the
two dominant galaxies computed with the 10-6 scale combination.
Additional locations are shown as in Figs.~\ref{fig:2310-8} and
\ref{fig:2310-7}. The two view directions of Fig.~\ref{fig:2310-63D}
are indicated. Black shaded areas show the diffuse light emission
maxima, with source 4 north of NGC~4874. The secondary X-ray peak from
Finoguenov (2004, private communication) is shown as open
circle on the right.}
\label{fig:2310-6}
\end{figure}

\begin{figure}
\caption[]{ R band images. Upper figure: view 1 of residual
image (10-6 combination scale) in Fig.~\ref{fig:2310-6}.  Lower
figure: view 2 of residual image (10-6 combination scale) in
Fig.~\ref{fig:2310-6}. Angles of view are shown in
Fig.~\ref{fig:2310-6}.  Additional locations of our sources and of
NGC~4889 are shown.}
\label{fig:2310-63D}
\end{figure}

The 10-8 scale combination shows a single roughly circular large scale
emission located north of NGC~4874, hereafter source~1 (see
Fig.~\ref{fig:2310-8} and Table 1). This source does not seem to be 
associated with any object but NGC~4874. There is however a
significant shift between NGC~4874 and the center of this emission. We
note that we do not detect any such emission close to NGC~4889. It is
therefore very unlikely that this source is a numerical artefact due
to the presence of a bright galaxy (NGC~4874) nearby.
Fig. \ref{fig:source1} shows maps of source 1 for the 10-6, 10-7 and
10-8 combination scales.

Using smaller scales (10-7), we find evidence for two other emissions
in addition to the previous source (see Fig.~\ref{fig:2310-7} and Table 1).
One of them (hereafter source 2) is located between NGC~4874 and
NGC~4889. This emission is centered $\sim$2 arcmin from the center of
the faint galaxy density map (Biviano et
al. 1996). Fig.~\ref{fig:source2} shows maps of source 2 for the 10-6,
10-7 and 10-8 combination scales.

The other emission (hereafter source 3) coincides exactly with the center of
the secondary X-ray emission (Finoguenov, private communication) and north
west of the dynamical center of the NGC~4874 group. 
Looking at the image of the object, we see that this
source is associated with the ``plume'' already reported by Gregg $\&$
West (1998). Fig. \ref{fig:source3} shows maps of source 3 for the
10-6, 10-7 and 10-8 combination scales.

Using scales 10-6, we detect several sources (Figs.~\ref{fig:2310-6}
and \ref{fig:2310-63D}). One is associated with source 2. This means
that source 2 has multiscale components.  This component of source 2
nearly coincides with the center of the cluster bulk X-ray emission (Neumann et
al. 2003).  We also detect a source north of NGC~4874 (hereafter
source 4). This source is possibly associated with a single galaxy
even if no clear sign of galaxy disruption appears in the R band
image. This source is only visible with the 10-6 combination scale
(see Fig.\ref{fig:source4} ).

Finally, several sources are detected north and south of source 3,
implying that source 3 has multiscale components. The southern part of
the small scale component of source 3 is located very close to the
dynamical center of the NGC~4874 group.

\subsection{Statistical significance of the detections of 
sources 1, 2, 3 and 4}

In order to estimate the statistical significance of the detections of the
four sources, we computed the signal-to-noise maps in the dominant
galaxy cluster region of the R band residual images (see
Fig.\ref{fig:stat}). The contours overplotted on these images show
respectively the 4 and 5$\sigma$ levels for the 10-6 scale
combination, the 3.5 and 4$\sigma$ levels for the 10-7 scale
combination and the 3 and 3.5$\sigma$ levels for the 10-8 scale
combination.  We clearly see that sources 2, 3 and 4 are detected
above the 4$\sigma$ level with the 10-6 scale combination. Using the
10-7 scale combination, sources 2 and 3 are detected above the
3.5$\sigma$ level. Source 1 is also detected above the 3$\sigma$ level
using the 10-8 scale combination.

Therefore these four sources appear to be significantly detected.

\subsection{Comparison with the literature}

To assess the quality of our treatment, we compared our
results with those of Gregg $\&$ West (1998). They have detected three
small scale sources of diffuse light and one large scale
source. Searching in our data in the same positions, we redetected
what we call source 3 (source 1 of Gregg $\&$ West: the largest scale
source they detected). This source is brighter in our
data because since we detect this source at a much larger scale than
Gregg $\&$ West (1998) we integrate more flux. The three other
sources of Gregg $\&$ West (1998) are not present in our residual
images. However, we were not expecting to detect them in the residual
images since we only search for large scale features. These three
additional sources have very small scales and even the 10-6 scale
combination removes them from the residual images.  This is confirmed
by the fact that these three sources are indeed visible in the final
object images.

For the same reason, we do not detect the same diffuse light
concentrations as Bernstein et al. (1995). Their field was directly
south of NGC~4874 and all their detected diffuse light was part of the
cD halo. At this scale, these sources have been removed from our
residual images.

\section{Results in the B and V bands}
\label{resultsBV}

\subsection{B and V bands}

Results of the analyses of the region of the dominant galaxies are
shown in Figs. \ref{fig:source1col}-\ref{fig:source3col} for the B, V
and R bands, and in Table 1. These figures show that source 1 is
detected in V and R but is not visible in B, while sources 2 and 3 are
detected in B, V and R, and source 4 is only visible in R.  Using  
combination scale 10-7 with V band data leads to
results similar to those obtained in the R band.

\subsection{Detection limits and colors of the various sources of
diffuse emission} 

We estimated the detection limits in the B, V and R bands assuming
that we were not able to detect large scale concentrations of diffuse
light exceeding the background level by less than 1$\%$. At this
level, the surface brightness detection limits are 25.4
mag/arcsec$^2$ in R, 25.8 mag/arcsec$^2$ in V and 27.2
mag/arcsec$^2$ in B.

In order to measure colors for all four sources, we used the
detections in the R band and extracted the fluxes inside the same
regions in the V and B bands. When the sources are not detected,
colors are therefore only limits.

Considering the values in Table 1, source 4 (with mean surface
brightnesses of R=25.1 mag/arcsec$^2$, V = 25.9 mag/arcsec$^2$ and
B=26.4 mag/arcsec$^2$) is detectable in R and in B, but not in
V. This explains why it is not visible in V. Since source 4 is
not visible in B either, it must be because this source is fainter in
B than in R (that is, it is intrinsically red).  Sources 1, 2 and 3 are
detectable in the B, V and R bands (but source 1 is not detected in B).

We can also note that, even if sources 1, 2 and 3 are detected in
several bands, the locations of the emission peaks are shifted from
one band to another, sometimes by a few arcminutes (see
Figs. 9-15). We also applied the same computation to source 4
even if no significant concentration of diffuse light is present in
the B and V band data. Color computations are summarized in Table
\ref{tab:col}.

Sources 3 and 4 exhibit colors close to those of elliptical/S0 galaxies.
Source 4 could be even redder, as the colors stated in Table 1 are only
lower values. Therefore, these sources probably have moderate star
forming activity. We note that
we have similar results as Gregg $\&$ West (1998) for source 3 given
our estimated error bars.

Despite the large error bars, values in Table 1 suggest that
source 2 has smaller B-R, more typical of a spiral-like than of
an elliptical galaxy. Source 1 could $possibly$ also have a spiral
like color, but as we only have lower values for B-R in Table 1, we
can not reach a definite conclusion for this source. 
This suggests a non-negligible star forming
activity at least inside source 2. Finally, source 2 appears bimodal or
elongated in the B band (detected with the 10-7 combination scale) but
not in the V and R bands, in agreement with the presence of extended
stellar formation activity due to several discrete galaxy disruptions.

\begin{table*}
\caption{Characteristics of sources 1, 2, 3 and 4 detected with the
10-7 combination scales. $\alpha$ and $\delta$ are
the J2000 central coordinates of the sources. Area is the (square) area
used to compute the magnitudes and is given in arcmin$^2$. The Gregg $\&$
West (1998, GW98) values are also reported for source 3.}
\begin{tabular}{llllllllll}
\hline
Source & $\alpha$ & $\delta$  & area & V & B-V  & B-V (GW98) & V-R &
V-R (GW98) & B-R  \\
\hline 
1   &  194.909 & 27.973 & 4.88 & 14.91$\pm$0.31 & $\geq$0.67 &  &
0.42 & & $\geq$1.09 \\
2   &  194.964 & 27.966 & 2.82 & 15.36$\pm$0.34 & 0.59 &  & 0.48 & &
1.07 \\
3   &  194.908 & 27.970 & 5.30 & 14.59$\pm$0.31 & 0.93 & 0.93 & 0.36 &
0.57 & 1.29 \\
4   &  194.910 & 28.015 & 2.78 & 15.95$\pm$0.35 & - &  &  $\geq$0.88  & &
$\geq$1.30 \\
\hline
\end{tabular}
\label{tab:col}
\end{table*}

\begin{figure}
\caption{Source 1 in the R band for the 10-6 (a), 10-7 (b) and 10-8 (c)
combination scales.}
\label{fig:source1}
\end{figure}

\begin{figure}

\caption{Source 2 in the R band for 10-6 (a), 10-7 (b) and 10-8 (c)
combination scales.}
\label{fig:source2}
\end{figure}

\begin{figure}

\caption{Source 3 in the R band for the 10-6 (a), 10-7 (b) and 10-8 (c)
combination scales.}
\label{fig:source3}
\end{figure}

\begin{figure}
\caption{Source 4 in the R band for the 10-6 (a), 10-7 (b) and 10-8 (c)
combination scales.}
\label{fig:source4}
\end{figure}

\begin{figure}
\caption{S/N maps for the 10-6 (upper left), 10-7 (upper right) and
10-8 (low center) scale combinations. Overlaid contours are for the
4-5 $\sigma$ levels (upper left),  3.5-4 $\sigma$ levels (upper right) and
3-3.5 $\sigma$ levels (low center).}
\label{fig:stat}
\end{figure}

\begin{figure}
\caption{Source 1 for the 10-7 combination scale in the B (a), V
(b) and R (c) bands.}
\label{fig:source1col}
\end{figure}

\begin{figure}
\caption{Source 2 for the 10-7 combination scale in the B (a), V
(b) and R (c) bands.}
\label{fig:source2col}
\end{figure}

\begin{figure}
\caption{Source 3 for the 10-7 combination scale in the B (a), V
(b) and R (c) bands.}
\label{fig:source3col}
\end{figure}

\section{Towards a general scenario for the Coma cluster evolution?}
\label{scenario}

We assume in the following that diffuse light is coming from disrupted
or harassed galaxies, a hypothesis which is supported by several
studies (e.g. Gregg $\&$ West 1998, Secker et al. 1998, Conselice $\&$
Gallagher 1999, Adami et al. 2000).

\subsection{Dynamics of the cluster}

As discussed below, our results can help us to understand better the
dynamics of the Coma cluster.

First, we do not detect any source of diffuse light around NGC~4889
while we have significant multi-scale sources around NGC~4874. This
might have been the case if NGC~4889 had been present in the
Coma cluster for a longer time than NGC~4874. NGC~4889 would then have
had enough time to reaccrete the diffuse material produced for example
by the disruption of a small galaxy or galaxy harassment. This is also
supported by the fact that NGC~4874 has a central velocity dispersion
higher than NGC~4889 (398 km/s against 275 km/s, Moore et al. 2002)
and is also brighter. This tends to favor the notion that
NGC~4889 has been in a dense environnement (the Coma cluster) for a longer 
time than NGC~4874.

Second, source 1 seems to be associated with NGC~4874 but is located north
of the galaxy. Let us assume that NGC~4874 and its surrounding group
are moving north through the Coma cluster (e.g. Donnelly et
al. 1999). We know that NGC~4874 will lose its kinetic energy much
more efficiently than all other galaxies in its group because of its large
mass (e.g. Sarazin 1986). Hence, most of the galaxies of the NGC~4874
group will, therefore, be ahead of NGC~4874 towards the north. This is
confirmed by the fact that the dynamical center of the NGC~4874 group
is also located north of NGC~4874 (Gurzadyan $\&$ Mazure 2001). We
know that tidal disruptions will occur preferentially in low mass and
late type galaxies. We also know that the NGC~4874 group has a quite
late type galaxy content (Gurzadyan $\&$ Mazure 2001). In this
scenario, material dispersed by disruption or galaxy harassment from
the NGC~4874 group will be located north of NGC~4874, as observed for
example with sources 3 and 4. Regarding source 3, one of the two
galaxies involved in this disruption is indeed attached to the
NGC~4874 group (Gurzadyan $\&$ Mazure 2001).  This motion towards the
north is also confirmed by recent X-ray data. Donnelly et al. (1999)
detected a hot region in the Coma cluster X-ray gas north of NGC~4874
using ASCA. This could be interpreted as a shock due to the motion of
NGC~4874 through the Intra Cluster Medium. This detection was not
confirmed in a first analysis of XMM data by Arnaud et
al. (2001). However, Neumann et al. (2003) detected with the same XMM
data a barely significant very elongated hot area located north of
NGC~4874. This could therefore be the hot front associated with the
motion of NGC~4874.

Source 2 has multiscale components and is located at equal distances
from the two dominant Coma cluster galaxies. This is not surprising because it
is the place where the tidal forces are the most intense in the
cluster, and therefore where galaxy disruptions would occur. It is
not clear, however, if this diffuse light source is associated with
the Coma cluster potential itself or if it lies here only
because of the influence of the two dominant galaxies.  This area
shows significant X-ray emission in excess of the global Coma
cluster emission (e.g. Biviano et al. 1996, Neumann et al. 2003,
Finoguenov et al. 2004 private communication) and this emission could
be associated with the true Coma X-ray center. Similarly, Biviano et
al. (1996) computed the position of the faint galaxy density map
center and associated this center (close to the X-ray center computed
by Neumann et al. 2003) to the true Coma cluster center.  However, the
diffuse light source is not located exactly at the same place as shown
in Figs.~\ref{fig:2310-7} and \ref{fig:2310-6}.
The explanation probably lies in the multiscale content
of source 2. The small scale component (detected with the 10-6 scale
combination) could be due to recent disruptions caused by the strong
tidal forces in this region while the largest component (detected 
with the 10-7 scale combination) could simply be the result of older
tidal disruptions retained at this place because it is close to the
bottom of the Coma cluster potential. This does not explain, however,
why the large scale component of source 2 is located significantly
north of the Coma cluster center (faint galaxy center and global X-ray
center).  Additional studies including numerical simulations are
needed in order to confirm or invalidate this scenario.

\subsection{The nature of the diffuse light}

Assuming that all this large scale diffuse light is coming from
totally disrupted galaxies, we computed the magnitude of a single
galaxy able to produce such an amount of diffuse light. For the 10-7
scale combination, we would need an R$\sim$13 or $M_R \sim -22$ galaxy
(or $\sim$$3\ 10^{11}\ M_\odot$ assuming the same M/L ratio as for the
Milky Way). This is a galaxy brighter than M$^{*}$ and is comparable
to the calculation of Gregg $\&$ West (1998), who estimated that an
$M_R=-22$ galaxy would produce the whole detected amount of small scale
diffuse light in the Coma cluster. This is not, however, our favorite
scenario as small galaxies are easier to disrupt than massive ones.

We also note that all this diffuse light represents only a negligible
fraction of the total Coma luminosity (e.g. Girardi et
al. 2002). However, if we limit ourselves to the cluster  core (where
the diffuse light is concentrated), the fraction becomes more
important. In the very center, it represents up to 20$\%$ of the
cluster galaxy luminosity. Assuming that
the central M/L ratio of Coma varies from 10 to 100 (e.g. Bernstein et
al. 1995), it lowers the M/L values by 15$\%$.

All these estimates assume that there is no significant ongoing star formation
in the material emitting the diffuse light.
Gregg $\&$ West (1998) computed a color for source 3 close to that of
an elliptical galaxy. This means that this diffuse light source is
probably not forming significant amounts of stars.  As mentionned above (See
Table 1 and Section 5.2), to
estimate the star formation rate in the various sources in which we
detected diffuse light emission, we used our data to compute colors.
To summarize the results: on the
one hand, we confirm the results of Gregg $\&$ West (1998, based on
HST observations) for source 3; source 4 is also probably a quiet
star forming region given its colors (probably even redder than the
lower values given in Table 1). On the other hand, source 2 (and
possibly source 1) has low B-R colors (blue), typical of a spiral galaxy
(e.g.  Karick et al. 2003). We note that this last result is only
marginally significant given the large uncertainties in Table 1.
The extent of source 2 in the B band is larger (and
bimodal) than in other bands. This suggests star forming activity in
the outskirts of this source.

To explain these results we propose the following
scenario: spiral galaxies are disrupted or harassed very efficiently
by the combined tidal forces of NGC~4874 and NGC~4889 at the location
of source 2 (e.g. Conselice et
al. 2003). This leads to an abundant infall of still star-forming
material in regions close to the bottom of the Coma cluster
gravitational potential. This potential probably confines the matter
at its present place. In agreement with this scenario, Stickel et
al. (1998), using ISO data, detected a significant amount of dust
located exactly where we detect source 2 while source 1 was less
evident in their data. This is in good agreement with the less
clear nature of source 1 given its marginally blue color (see Table
1). They concluded that a total amount of dust
between $6.2\ 10^7$ and $1.6\ 10^9\ M_\odot$ would explain their
observations. Assuming a dust to gas ratio between $1.3\ 10^{-5}$ and
$3.2\ 10^{-4}$ (Stickel et al. 1998), we would predict a gas mass of
between $1.9\ 10^{11}$ and $1.2\ 10^{14}\ M_\odot$. This is close to
our estimate of the equivalent of one bright galaxy needed to generate
the detected diffuse light.  Moreover, Stickel et al. (1998) concluded
that since dust is destroyed continuously by the hot ICM, the dust
detected at the location of source 2 would be due to quite recent
galaxy stripping. This is in good agreement with the hypothesis 
that source 2 has a recent component 
(appearing only in the 10-6 combination
scale) originating from recent tidal disruptions or galaxy stripping
due to the intense tidal forces between the two dominant
galaxies. This would occur easily even for late type galaxies,
generating diffuse light with a low B-R color (blue), as observed in our
data.

\section{Conclusions}
\label{conclusions}

By applying a multi scale wavelet analysis and reconstruction
technique to large images of the Coma cluster in various bands, we
have shown the existence of several large scale sources of diffuse
emission. The
quality of the results obtained shows the power of this method
for detecting faint diffuse emission in a rather crowded field.

No source of diffuse emission is detected around NGC~4889, possibly 
suggesting
that this galaxy is older and has had time to accrete any diffuse material
that may have been available in its surroundings.

On the other hand, we detected four diffuse sources around NGC~4874. 
Sources~3 and 4 have
colors typical of early type galaxies, while source 2
probably shows signs of
star formation activity, mainly in the outskirts of the concentration.

We suggest that a possible explanation for the
origin of diffuse emission is that between
the two main galaxies NGC~4874 and NGC~4889, spiral galaxies have
recently been disrupted, and matter still forming stars has
accumulated at the bottom of the cluster potential well. Other sources
with colors typical of elliptical galaxies have probably been
generated by older disruptions or disruptions of early type galaxies.

The characteristics of the sources of diffuse emission agree with a
scenario in which the group around NGC~4874 is moving northwards, as
suggested by recent X-ray data.

\begin{acknowledgements}

The authors thank the referee for useful and constructive comments, J.
Katgert for english corrections, 
M. Trayna for efficient numerical support and Olivier Ilbert for
useful discussions. 

\end{acknowledgements}

\end{document}